\documentclass[10pt]{webofc}
\usepackage[varg]{txfonts}   
\usepackage{multirow}
\usepackage{array, makecell}

\begin{document}

\title{WLCG Authorisation}
\subtitle{from X.509 to Tokens}

\author{
        \firstname{Brian} \lastname{Bockelman}\inst{6} \and
        \firstname{Andrea} \lastname{Ceccanti}\inst{3} \and
        \firstname{Ian} \lastname{Collier}\inst{4} \and
        \firstname{Linda} \lastname{Cornwall}\inst{4} \and
        \firstname{Thomas} \lastname{Dack}\inst{4} \and
        \firstname{Jaroslav} \lastname{Guenther}\inst{1} \and
        \firstname{Mario} \lastname{Lassnig}\inst{1} \and
        \firstname{Maarten} \lastname{Litmaath}\inst{1} \and
        \firstname{Paul} \lastname{Millar}\inst{2} \and
        \firstname{Mischa} \lastname{Sallé}\inst{5} \and
        \firstname{Hannah} \lastname{Short}\inst{1}\fnsep\thanks{\email{hannah.short@cern.ch}} \and
        \firstname{Jeny} \lastname{Teheran}\inst{7} \and
        \firstname{Romain} \lastname{Wartel}\inst{1} 
}

\institute{European Organization for Nuclear Research (CERN) \and 
Deutsches Elektronen-Synchrotron, Germany (DESY) \and 
Istituto Nazionale di Fisica Nucleare, Italy (INFN) \and
Science and Technology Facilities Council (UKRI-STFC), United Kingdom \and
Nationaal Instituut voor Subatomaire Fysica (Nikhef), Netherlands \and
Morgridge Institute for Research, United States \and
Fermi National Accelerator Laboratory, United States 
}

\abstract{%
  The WLCG Authorisation Working Group was formed in July 2017 with the objective to understand and meet the needs of a future-looking Authentication and Authorisation Infrastructure (AAI) for WLCG experiments. Much has changed since the early 2000s when X.509 certificates presented the most suitable choice for authorisation within the grid; progress in token based authorisation and identity federation has provided an interesting alternative with notable advantages in usability and compatibility with external (commercial) partners. The need for interoperability in this new model is paramount as infrastructures and research communities become increasingly interdependent. Over the past two years, the working group has made significant steps towards identifying a system to meet the technical needs highlighted by the community during staged requirements gathering activities. Enhancement work has been possible thanks to externally funded projects, allowing existing AAI solutions to be adapted to our needs. A cornerstone of the infrastructure is the reliance on a common token schema in line with evolving standards and best practices, allowing for maximum compatibility and easy cooperation with peer infrastructures and services. We present the work of the group and an analysis of the anticipated changes in authorisation model by moving from X.509 to token based authorisation. A concrete example of token integration in Rucio is presented.
}

\maketitle
\section{Introduction}
\label{intro}
This paper describes ongoing work by the WLCG Authorisation Working Group \cite{twiki} to transition the Worldwide LHC Computing Grid's authorisation model from X.509 Certificates \cite{x509} and certificate proxies \cite{rfcproxy} to tokens. The tokens are defined as JSON Web Tokens (JWT) \cite{jwt}, to be provisioned over OpenID Connect (OIDC) \cite{oidc} and OAuth2 \cite{oauth2} workflows. 

\subsection{Motivation}
\label{motivation}
When X.509 was chosen in the early 2000s, there was no mature alternative. The Globus Toolkit \cite{globus}, chosen by WLCG, supported X.509 and provided a functional solution for distributed authentication and authorisation, when coupled with policies controlled by the Interoperable Global Trust Federation (IGTF) \cite{igtf}. Despite providing a solution, certificate management proved a steep learning curve for researchers and the authorisation model adopted (although sufficient at the time) is outdated by today's standards regarding the protection of privacy of user data. 

In the meantime, alternative authentication and authorisation technologies have emerged and X.509 is no longer considered the most suitable option. It is increasingly common that many off-the-shelf software choices and commercially hosted services offer inbuilt options for federated authentication, using standards such as SAML or OIDC. Integrating grid software with such third party systems has therefore become progressively more difficult, since a lot of X.509 developments are deeply integrated in the grid software stack. Users are now well accustomed to web-based authorisation workflows that use OAuth2 or OIDC to delegate access rights, meaning that adoption of such technologies would present a more user-friendly experience. These factors came together to build a strong motivation for redesigning the WLCG authorisation model based on JWTs over OIDC and OAuth2.

\subsection{WLCG Authorisation Working Group}
\label{wg}

The WLCG Authorisation Working Group was formed in the summer of 2017, at a time when multiple activities were independently beginning to seriously consider token based authorisation. SciTokens \cite{scitokens} had produced a working system in the US focusing on capability-based authorization and an increasing number of efforts were ongoing in Europe to enable token based workflows, in particular in the context of the INDIGO DataCloud project \cite{indigo} and EGI \cite{egi}. Experts from multiple domains came together in this group to chart a path towards token based authorisation for WLCG. Work to enhance software was supported by several European Commission Projects: EOSC-Hub \cite{eoschub}, EOSC Pilot \cite{eoscpilot} and AARC2 \cite{aarc}. The group's objective was to understand and meet the requirements of a future-looking authorisation service for WLCG experiments. In particular, effort focused on addressing usability, i.e. removing the need for users to manage certificates themselves and enable more intuitive authentication methods. A second objective was to simplify integration with third party services and software by adopting industry standards. A significant effort has been made to be as inclusive as possible, by including resource and service providers during the construction of the schema and the testing phase, as well as making all outputs available to the wider community such as through the FIM4R \cite{fim4r} workshop at Fermilab in September 2019. Documentation and a list of meetings are maintained on the WLCG Authorisation Working Group Twiki \cite{twiki}.   

\section{Milestones}
\label{milestones}
\begin{figure}[h]
\centering
\includegraphics[width=12cm,clip]{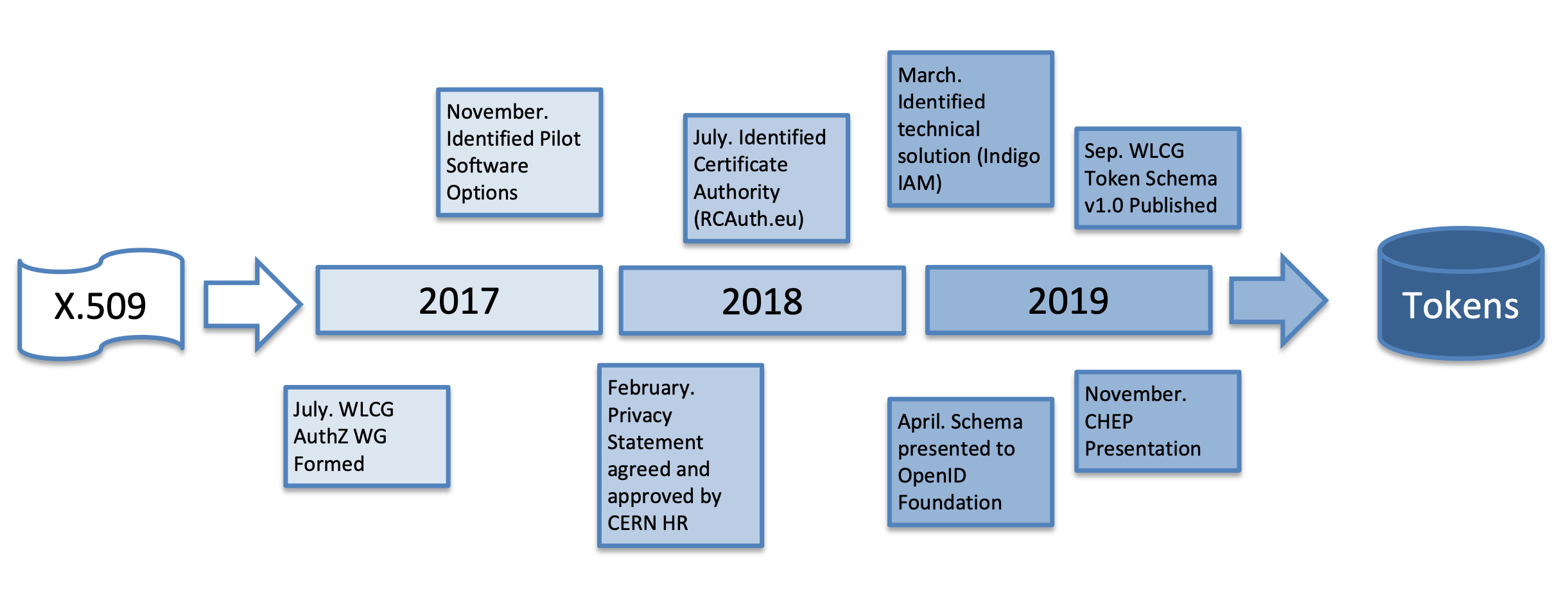}
\caption{Key milestones achieved between 2017 and 2019 towards a token based authorisation scheme. Note, it is not implied that the final goal has been reached; much additional effort is required to fully transition from X.509 to tokens.}
\label{milestones-fig}
\end{figure}

Several key milestones have been achieved since the WLCG Authorisation Working Group was formed, with major progress included in Figure \ref{milestones-fig}. In particular, a competitive study of two pilot solutions was made, resulting in an identified technical solution in March 2019. A schema was negotiated between multiple stakeholders over the course of a year and was published on Zenodo \cite{zenodo} in September 2019. 

\section{Technical Design}
\label{design}

\begin{figure}[h]
\centering
\includegraphics[width=10cm,clip]{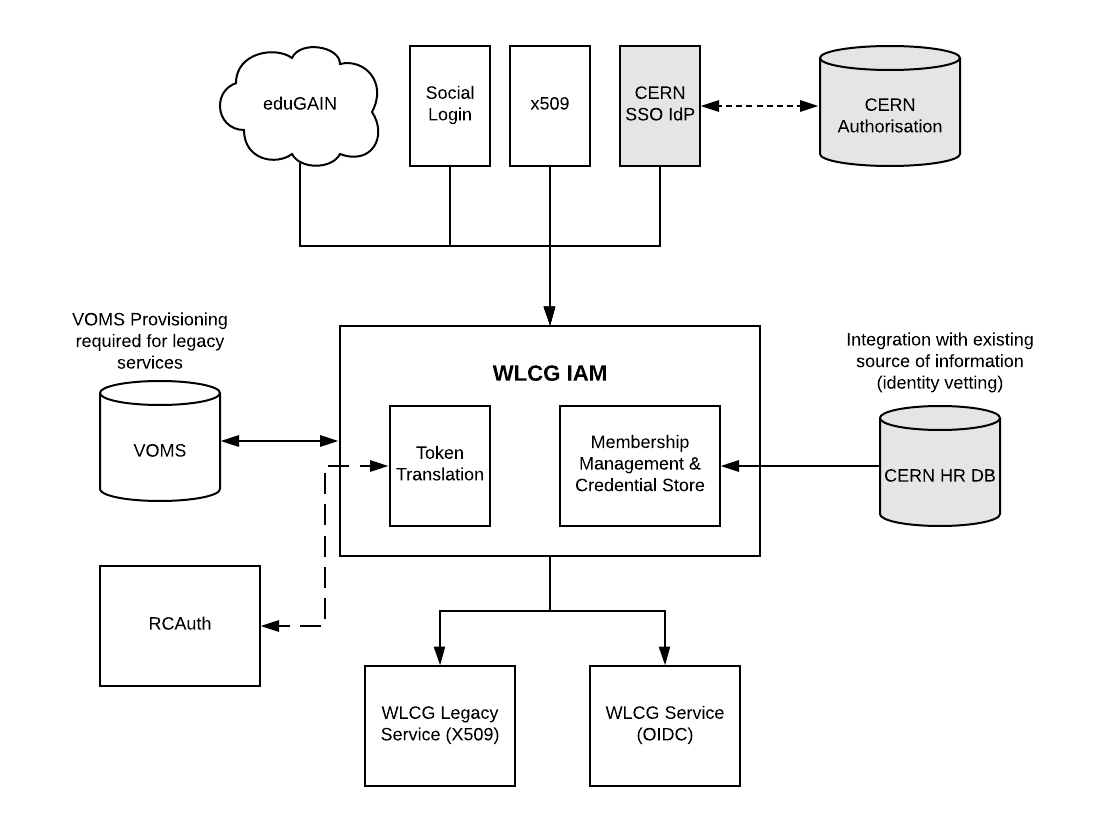}
\caption{Design of the WLCG Token Authentication and Authorisation Infrastructure. Grey boxes represent configurable components, in this case tuned for the CERN environment.}
\label{design-fig}
\end{figure}

The INDIGO IAM \cite{iam} software was chosen as the core of WLCG's future, token based authentication and authorisation infrastructure. In Figure \ref{design-fig}, we see the WLCG IAM (a WLCG Instance of INDIGO IAM) in the centre, proxying multiple authentication channels, centralising authorisation, providing token translation to and from X.509, and serving as a uniform token issuer for downstream WLCG Services. The Token Translation is performed using the RCauth.eu \cite{rcauth} online Certificate Authority and users can request a certificate from the WLCG IAM User Interface. This architecture is in line with guidelines from the AARC Project, including the AARC Blueprint Architecture.

\subsection{Usage Flow}

\begin{figure}[h]
\centering
\includegraphics[width=9cm,clip]{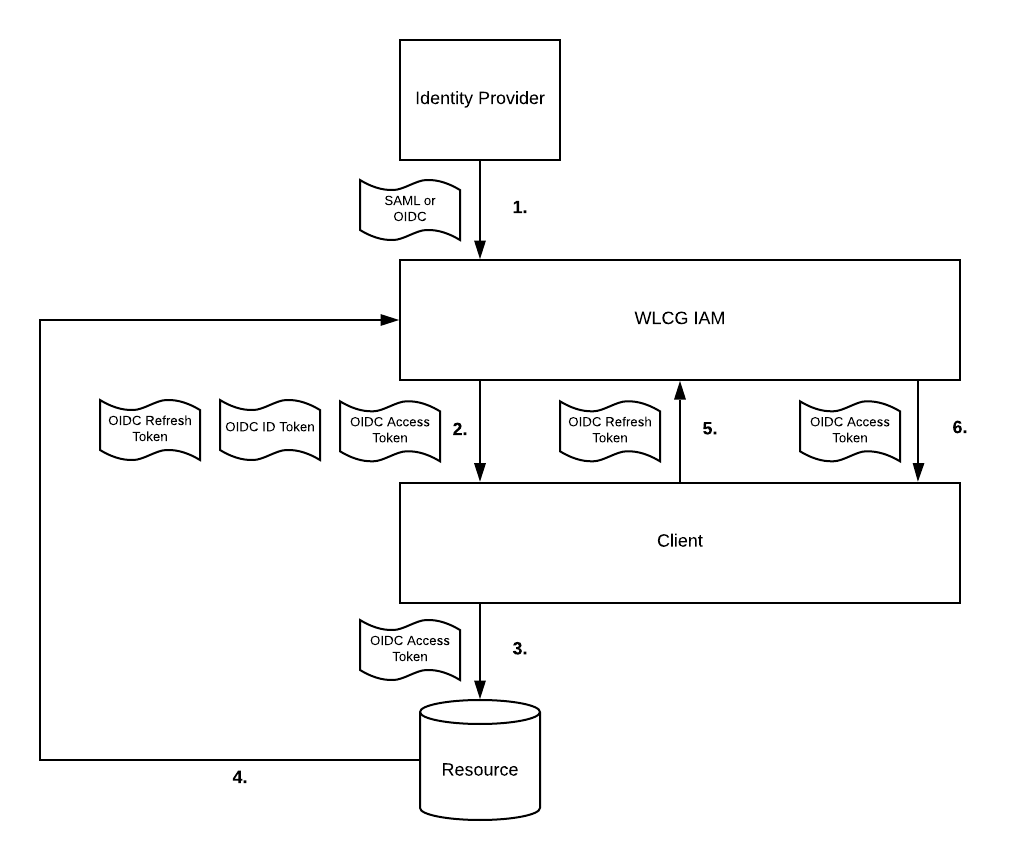}
\caption{Anticipated token flow, starting from a SAML or OIDC token being issued from the Identity provider to WLCG IAM. WLCG IAM then issues multiple OIDC tokens to downstream clients, granting access to protected resources.}
\label{flow-fig}
\end{figure}

It is envisaged that there will be a small number of registered clients for WLCG and a much larger number of unregistered Resources. The user's authentication from their Identity Provider is technically decoupled from the tokens issued by the WLCG IAM. A user may authenticate with SAML \cite{saml}, OAuth2, OIDC, X.509 or potentially any future protocol that is accepted by WLCG IAM. A more detailed description of IAM~\cite{beyond-x509} and the work done to address WLCG requirements is given in a dedicated contribution to this journal~\cite{beyond-x509-ii}. Figure \ref{flow-fig} shows the anticipated token workflow, the numbered steps are explained below: 
\begin{enumerate}
    \item A SAML or OIDC credential is sent from an Identity Provider to WLCG IAM
    \item WLCG IAM sends three OIDC tokens to a registered Client, the ID Token, Access Token and Refresh Token
    \item The Access Token is used to authorise access to downstream Resources (such as storage)
    \item The Resource validates the token against known trust roots. The trust roots may be cached in advance to avoid a high number of round trips to the token issuer.
    \item When the Access Token expires, the Client may use the Refresh Token to request a newly valid Access Token from WLCG IAM 
    \item WLCG IAM returns an Access Token to the Client, which may repeat steps 3 to 6
\end{enumerate}

\section{Schema}

The WLCG Common JWT Profiles document was published on September 25th, 2019 \cite{schema}. This document describes how WLCG users may use the available geographically distributed resources without X.509 credentials.  In this model, clients are issued with bearer tokens; these tokens are subsequently used to interact with resources.  The tokens may contain authorization groups and/or capabilities, according to the preference of the Virtual Organisation (VO), applications and relying parties. 

Wherever possible, the document builds on existing standards when describing profiles to support current and anticipated WLCG usage.  In particular, three major technologies are identified as providing the basis for this system: OAuth2 (RFC 6749 and RFC 6750), OIDC  and JWTs (RFC 7519). Additionally, trust roots are established via, among others, the OpenID Discovery \cite{oidcdisco} mechanism or OAuth2 Authorization Server Metadata (RFC 8414 \cite{oauth2metadata}). The document provides a profile for OAuth2 Access Tokens and OIDC ID Tokens, the claims for which are shown in Table \ref{table:1}. Although many WLCG requirements were covered by standard claims defined in RFC7519 \cite{jwt} or OIDC core \cite{oidc}, a claim to convey assurance was taken from the Research and Education Federations (REFEDS) \cite{refeds} specification. Claims to convey version (\textit{wlcg.ver}) and groups (\textit{wlcg.groups}) have been defined within the WLCG schema document itself.

\begin{table}[h]
\centering
\begin{tabular}{ p{3.5cm}|p{3cm}|p{2.5cm}|p{2.5cm} }
Claim & Origin & WLCG OIDC ID Token & WLCG OAuth2 Access Token \\ \hline \hline
sub & RFC7519 & X & X \\\hline
exp & RFC7519 & X & X \\\hline
iss & RFC7519 & X & X \\\hline
acr & OIDC core & X & X \\\hline
aud & RFC7519 & X & X \\\hline
iat & RFC7519 & X & X \\\hline
nbf & RFC7519 & X & X \\\hline
jti & RFC7519 & X & X \\\hline
eduperson\_assurance & REFEDS & X & X \\\hline
wlcg.ver & WLCG  & X & X \\\hline
wlcg.groups & WLCG  & X & X \\\hline
auth\_time & OIDC core  & X & \\\hline
standard OIDC claims & OIDC core  & X \\\hline
scope & \textit{Inspired by Auth token exchange draft} & & X \\\hline
\end{tabular}
\caption{Token Claims used in the WLCG Schema, including their origin in either RFC7519 \cite{jwt}, OIDC core \cite{oidc}, the Research and Education Federations (REFEDS) \cite{refeds} or the WLCG Schema itself \cite{schema}.``X'' indicates their use in ID and Access tokens. Claim definitions can be found at the origin and are not included here for brevity.}
\label{table:1}
\end{table}

\subsection{Authorisation}
Two forms of Authorisation exist in parallel in the WLCG Common JWT Profile Schema; Group based Authorisation and Capability based Authorisation. Groups are semantically equivalent to existing VOMS groups, i.e. a group contains a list of users. A capability is the ability to perform an action, optionally at a specific path, e.g. \textit{"the bearer of this token is authorised to write to /data"}. Extended examples are included in the Schema document \cite{schema}.

\section{WLCG Token Issuer}
\label{infn-iss}
In late 2019, a WLCG token issuer was set up at the Istituto Nazionale di Fisica Nucleare (INFN) to provide a stable platform against which software enhancement could be tested. The token issuer was made available at \url{https://wlcg.cloud.cnaf.infn.it/}, primarily to facilitate tests planned by the WLCG DOMA Working Group. Clients are able to register to receive authentication tokens from the issuer over OIDC, and a web based platform is available to visualise tokens and their content. A VOMS attribute authority linked to this IAM instance has also been deployed, to allow users to get VOMS attribute certificates for the \texttt{wlcg} VO. The objective is to demonstrate IAM interoperability with the current X.509 WLCG AAI and enable a gradual transition to tokens.

\section{WLCG Tokens working example: Rucio}

Rucio \cite{rucio} is one of the main Data Management tools used by WLCG experiments and is a key service to be enhanced to accept WLCG tokens. During 2019, the Rucio authentication and authorization mechanism was extended to support JWTs using the Open ID Connect protocol (following the OAuth 2.0 specifications). The implementation is based on the OIDC certified \cite{oidccert} \textit{pyoidc} \cite{pyoidc} library and follows the WLCG specification. During this process several improvements to the schema were identified, which will be incorporated into the next version of the WLCG token profile specification. To perform operations with Rucio \textit{1.22.0.dev3} a user can now log in via the authorization code flow, which has been implemented in both the Rucio WebUI and Rucio command line interface (CLI). Several CLI login workflows are supported - including username and password, or session validation using a browser - to fulfil a variety of researcher use cases. The Rucio REST API has been configured as an OAuth2 protected resource and accepts tokens from trusted OIDC providers (including the WLCG token issuer discussed in Section \ref{infn-iss}).

\subsection{Rucio and Downstream Storage}
Tests have been successfully performed to demonstrate token based authorization from Rucio to the File Transfer Service (FTS) \cite{fts}. When a Rucio user requests a file transfer, Rucio can use either a JWT issued for the user or a JWT issued to the Rucio Admin user itself, which submits the job on the user's behalf. In order to allow fine grained access control downstream (Rucio $\rightarrow$ FTS3), token exchange and token refresh grant flows were also implemented. The first functional tests of a third party copy were performed (Rucio $\rightarrow$ FTS3 $\rightarrow$ dCache) and a version \textit{1.22.0.dev3} of Rucio supporting all these new features is currently deployed on the DOMA instance \cite{ruciodoma}. Additional tests with other downstream storage elements are planned.

\section{Conclusion}
The WLCG Authorisation Working Group is well on the way to enabling token based authorisation for WLCG workflows. This work has been possible thanks to the participation of multiple individuals and projects, who together have been able to identify a technical solution and define a token schema. Requirements from Virtual Organizations and other research infrastructures outside of WLCG have been included as a continuous cooperative process to guarantee interoperability. Preliminary token based workflows have been tested with many more planned for 2020. Further efforts are required to complete the enhancement of middleware, and to ensure that the existing WLCG policies are adhered to in the years to come.

\section*{}
\small{This work was achieved thanks to support from the following European Commission funded projects; EOSC-Hub, EOSC Pilot and AARC2. EOSC-hub receives funding from the European Union’s Horizon 2020 research and innovation programme under grant agreement No. 777536. EOSCPilot.eu has received funding from the European Commission’s Horizon 2020 research and innovation programme under the Grant Agreement no 739563. AARC2 is funded by the European Union’s Horizon 2020 research and innovation programme under grant agreement number 730941.  This material is based upon work supported by the National Science Foundation under Grant No. 1836650.}

\end{document}